\documentclass[conference]{IEEEtran}
\IEEEoverridecommandlockouts
\usepackage{latexsym,amssymb,amsfonts,graphics,indentfirst,CJK,CJKnumb}
\usepackage[cmex10]{amsmath}
\usepackage{clrscode}
   \usepackage[pdftex]{graphicx}
   \usepackage{cite}
  \DeclareGraphicsExtensions{.pdf,.jpg,.png}

\usepackage{multirow,array}
\title{ Evaluation on  the Financial Competitiveness of Chinese  Listed Real Estate  Companies \\ Based on Entropy Method
}
\author{\IEEEauthorblockN{Lin Wei}\thanks{Supported by Business Information Analysis Laboratory of Hangzhou.}
\IEEEauthorblockA{Business School of Zhejiang University City College\\
Hangzhou, China\\linw@zucc.edu.cn}
\and
\IEEEauthorblockN{Linbo Shao}\IEEEauthorblockA{School of Electrical Engineering and Computer Science\\ Peking University, Beijing 100871, China\\ shaolb@pku.edu.cn }
}

\begin{document}

\maketitle
\begin{abstract}
 The real estate is a pillar industry of China's national economy. Due to changes in policy and market conditions, the real estate companies are facing greater pressures to survive in a competitive environment. They must improve their financial competitiveness. Based on the conceptual framework of financial competitiveness, this paper presented a financial competitiveness evaluation index system, covering four aspects, including profitability, solvency, sustainable development and operational capacity.  Entropy value method is applied to determine the index weight. 105 listed real estate company's financial competitiveness are evaluated, the results show that: high-scoring company has strong profitability, sustainable development and operational capacity; low-scoring company has weak profitability and poor ability of sustainable development; solvency doesn't affect the company's financial competitiveness obviously.

 Key words:  Financial Competitiveness,  Evaluation, Entropy, Listed Real Estate  Companies
\end{abstract}

\section{Introduction}
During the past decade, Chinese real estate industry is developing rapidly. It has become an important pillar industry of national economy, and the rapidly rising prices of real estate have drawn concerns of the whole society. In 2010, the government departments issued a series of regulatory policies to ensure that the prices were under control. In 2011, the State Council promulgated the ``New National 8 Regulatory Policies ", and more stringent policies were implemented. At present, the cumulative effects of several rounds of regulation begin to show. Restrictions on real estate market transactions, has made great pressure of sale on the real estate companies. The companies face the challenges of that inventory turnover slows down, that cash flow continues tight, and that capital costs and financial risk increase. In order to survive and develop, real estate companies must improve their competitiveness. Real estate investment is capital-intensive investment, and the debt ratios of such companies generally get higher. The financing capacities of companies have significant impact on the development of enterprise. Therefore, for the real estate companies, financial competitiveness is an important aspect of the competitiveness of enterprises.

\section{Literature Review}
Since Stephen H. Hymer's first mention of the competitiveness of enterprises in his PhD Dissertation ``The International Operations of National Firms: A Study of Direct Foreign Investment" in 1960, the competitiveness of enterprises has become the focal point where both business and academia pay high attention to. However, the definition of competitiveness of enterprises has failed to reach an agreement so far.
At World Economic Forum (1994), the competitiveness of enterprises is defined as ``a company produces more wealth equilibrium balanced in the world market than the other \mbox{competitors."\cite{WEF}} In 1990 with their article titled ``The Core Competence of the Corporation", Prahalad and Hamel illustrated that Core Competencies are the collective learning in the organization, especially how to coordinate diverse production skills and integrate multiple streams of technologies. If core competence is about harmonizing strems of technology, it is also about the organization of work and the delivery of value.\cite{PCK} Capability theory regarded that the competitiveness of enterprises is a capability system, to accumulate, maintain and develop products and markets are the decisive factor of long-term competitive advantage. Differences in the ability of the enterprise is the source of sustainable competitive advantage.

Researches on financial competitiveness are still in the initial exploration stage. The meaning of the definition of financial competitiveness, is mainly from the capability theory view of the competitiveness of enterprises.
Wang Yanhui and Guo Xiaoming think that financial competitiveness is the demonstrated ability presented in the process of financial to achieve its business objectives, and it is acomprehensive strength resulting from the combination of financial strategy, financial resources, financial capacity, financial performance and financial innovation.\cite{wangguo}

Zhang Youtang and Fen Ziqin define the financial competitiveness as three dimensions including: the competitiveness of the financial strategy of adaptation to the environment; financial resources configuration competitiveness; financial interests together competitive. They suggest that the financial competitiveness is the value creation process under the \mbox{enterprise} strategic direction.\cite{zhang2008}

Cheng Yan thinks that financial competitiveness is a kind of integration capability to help corporate achieve sustainable competitive advantages, it is rooted in the ability of the financial system specifically for corporate finance. The core of the competitiveness is knowledge and innovation.\cite{cheng2008}
Cheng Yan hold that financial capability is a resource that can be applied to the force of financial control, including financial operating capacity, financial management capacity and financial adaptability, and it is the organic composition of the core competitiveness. \cite{cheng2008}

Hao Chenglin thinks that financial competitiveness is based on the value chain or supply chain enterprises group's capital investment and revenue generation activities and the financial relationship, with market competition as the driving force, around for competitive advantage, to the ability to create value for customers.\cite{hao1}

There are some studies about the evaluation of \mbox{financial} competitiveness. The existing literature mainly considers through selecting firms' financial index. Shen Airong established a comprehensive evaluation system of the financial competitiveness based on factor analysis, considering the profitability capability, debt paying capability, growth capacity and operation capacity with 13 selected financial \mbox{indicators.\cite{shen1}}

Feng Ran and Xiaoling Zhang starts from setting up a coherent conceptual and analytical framework covering different aspects, including profitability capability, debt paying capability, and operation capability. They presented an evaluation system based on 10 financial indicators. \cite{feng1}

Taking 22 real estate listed companies as sample,Wang Dongmei and Sun Zaoliang useing the factorial analysis method make ananalysis on their  financial competitiveness. The results show that the operational capability has the greatest impact on the company's financial competitiveness; profitability has a larger impact on the of the company's financial competitiveness; solvency has little impact on the company's financial competitiveness.\cite{wangzao1}

\section{Integrated Index Financial Competitiveness Evaluation System }
\subsection{Indicator selection}
According to the existing literatures, taking into account the availability of data, the financial competitiveness is decomposed into four parts including: profitability, solvency, sustainable development and operational capacity. The index system were designed as shown in Table \ref{table:ind}. Taking into account the characteristics of the real estate industry, capital intensity indicators are added to the operational capability section. All the data of indicators in Table \ref{table:ind} are obtained and calculated from the CSMAR database of Shenzhen Guo Tai'an information technology Co.Ltd.
\begin{table}[!ht]
  \centering
  \caption{The Financial Competitiveness Indicators}\label{table:ind}
  \begin{tabular}{|c|c|}
    \hline
    Category & Indicator \\
    \hline
     \multirow{3}{2cm}{Profitability capability} & Operating profit ratio \\
      \cline{2-2}
      & Return on assets \\
      \cline{2-2}
      & Return on invested capital \\
    \hline
    \multirow{4}{2cm}{Solvency} & Debt coverage ratio \\
        \cline{2-2}
		& Current ratio \\
        \cline{2-2}
		& Operating cash flow to operating profit ratio \\
        \cline{2-2}
		& Debt asset ratio \\
    \hline
    \multirow{5}{2cm}{Capacity for sustainable development}
    &Sustainable growth rate \\
    \cline{2-2}
	&Hedging and proliferating ratios \\
    \cline{2-2}
	&Total assets growth rate \\
    \cline{2-2}
	&Revenue growth rate \\
    \cline{2-2}
	&Net profit growth rate \\
    \hline
    \multirow{5}{2cm}{Operation capacity}
    &Receivables turnover \\
    \cline{2-2}
	&Inventory turnover \\
    \cline{2-2}
	&Total assets turnover \\
    \cline{2-2}
	&Rate of cost-profit \\
    \cline{2-2}
	&Capital intensity \\
    \hline
  \end{tabular}
\end{table}
\subsection{The principle of entropy and the steps to determine the index weight}
Entropy is a concept in thermodynamics. It was first proposed by Rudolf Clausius in 1850. In1877, Boltzmann linked the entropy S with the thermodynamic probability $\Omega$(that is, the number of microscopic quantum states), by
\begin{equation*}
    S = k \ln \Omega
\end{equation*}
where k is the Boltzmann's constant.

The concept of entropy in information theory was proposed by C.E. Shannon in 1948, which is ``a measure of information, choice and uncertainty".  He proved that the entropy H is of the form,
\begin{equation}\label{equ:shannon}
    H = - K \sum_{i=1}^{n} p_i \log p_i
\end{equation}

Where $K$ is a positive constant and  $\{p_1,p_2,\ldots,p_n\}$ is the probabilities of a set of possible events.

In information theory, entropy is a measure of uncertainty. The greater the amount of information, or the smaller the uncertainty, the smaller the entropy is; the smaller the amount of information, or the greater the uncertainty, the greater the entropy is. According to the characteristics of entropy, we can use entropy to determine the degree of dispersion of a indicator, the greater the entropy , the greater the degree of dispersion indicators, the greater the impact on the comprehensive evaluation,the greater the weight of the indicator.

According to the basic principle of the entropy method, the evaluation of the financial competitiveness as follows:

\emph{1.  Non-dimensionalization of indicators.}

The evaluation indicators are in different units of measurement; thus, the use of varied units may lead to inconsistent in evaluation. In order to eliminate the negative effects of using different units, every indicator would be non-dimensionalized.
Non-dimensionalization also considers the direction of each indicator. If an indicator is positive, or that the higher this indicator, the better such a company is, the non-dimensionalized data of indicator $j$ for the company $i$, denoted as $s_{ij}$, is computed as,
\begin{equation}\label{equ:one}
     s_{i\,j}=\frac{r_{i\,j} -\min\limits_{1 \le k \le m}{r_{i\,k}} }
     {\max\limits_{1 \le k \le m}{r_{i\,k}}-\min\limits_{1 \le k \le m}{r_{i\,k}}} \quad\,
\end{equation}

where $r_{ij}$ is the $j$th original indicator of the $i$th company, m is the number of indicators, and n is the number of companies. If such an indicator is inverse, i.e. the lower this indicator, the better such a company is, the non-dimensionalized data $s_{ij}$ is computed as,

\begin{equation}\label{equ:two}
        s_{i\,j}=\frac{\max \limits_{1 \le k \le m} r_{i\,k} - r_{i\,j}}{\max\limits_{1 \le k \le m}{r_{i\,k}}-\min\limits_{1 \le k \le m}{r_{i\,k}}} \quad\,
\end{equation}

\emph{2.  Estimation of Cumulative Distribution Function.}

Let $R_{j}$  be the set of data of indicator j for all companies, the distribution of indicator j is estimated by $R_{j}$, using the Kernel Density Estimation. The distribution of indicator j is illustrated by its cumulative distribution function (CDF)$\varphi_j(x)$. The CDF enjoys the property that it is a monotonically increasing function, and,
\begin{equation*}
    0\leq \varphi_j(x) \leq 1
\end{equation*}

\emph{3.  Computation of Entropy}

As defined in formula (\ref{equ:shannon}) the entropy is computed by the probabilities of the possible events. Since the indicators in our problem are not discrete but continuous, we proposed a new form of entropy. The probability in \mbox{equation \ref{equ:shannon}} is replaced by the value cumulative distribution function as mentioned above and the sum is replaced by the Integration. The continuous form of entropy for indicator j is defined as,
\begin{equation}\label{equ:three}
    H_j = -\mathbb{e} \int_0^1 \varphi_j(x) \ln \varphi_j(x) \textrm{dx}
\end{equation}
where the constant e is the base of the natural logarithms.

\emph{4.  Computation of the weight for each indicator}
   The weight for indicator $j$, $w_j$ is given by,
\begin{equation}\label{equ:four}
    w_j = \frac{H_j}{\sum_{k=1}^{m}{H_k}}
\end{equation}
where $H_j$ is the entropy of indicator $j$, $m$ is the number of indicators.

\emph{5.  Evaluation of Financial Competitiveness}

The Financial Competitiveness of each computer is evaluated by the integrated score. The integrated score for company $i$ is the function of its non-dimensionalized indicators $s_{ij}$ ,and weighted by $w_j$, denoted as,
\begin{equation}\label{equ:five}
    F_i = \sum_{j=1}^{m} w_j s_{ij} \;\quad i = 1,2,...,n
\end{equation}

\section{Empirical Results and Analysis}
\subsection{Sample Selection}
In accordance with the Commission's industry classification, we select the 108 listed real estate companies of A-share market in Shanghai and Shenzhen from 2010, three missing data samples are removed, and finally get the valid sample of 105. In this paper, all data are from the CSMAR database.

\subsection{Determining the Weights}
First, non-dimensionalize the sample data by formula (\ref{equ:one}) and (\ref{equ:two}). In the index system, the capital intensity and asset-liability ratio are inverse indicators (indicator value, the smaller the better),the others are positive indicators (indicator value, the bigger the better), all the dimensionless value of indicators will be between 0 and 1. Then the cumulative density function of each indicator $\varphi_j(x)$  is generated with kernel density estimation by MatLab. According to the formula (\ref{equ:three}) the entropy value of each indicator is calculated, and each weight of the indicator is calculated according to formula (\ref{equ:four}), as shown in Table \ref{table:enteopy}.

\newcommand{\PreserveBackslash}[1]{\let\temp=\\#1\let\\=\temp}
\newcolumntype{C}[1]{>{\PreserveBackslash\centering}p{#1}}
\newcolumntype{R}[1]{>{\PreserveBackslash\raggedleft}p{#1}}
\newcolumntype{L}[1]{>{\PreserveBackslash\raggedright}p{#1}}
\newcommand{\tabincell}[2]{\begin{tabular}{@{}#1@{}}#2\end{tabular}}
\newcommand{\tc}[1]{\multicolumn{1}{c|}{#1}} 
\begin{table}[!ht]
  \centering
  \caption{Entropy and Weight of Indicators}\label{table:enteopy}
  \begin{tabular}{|c|p{3.3cm}|r|r|}
    \hline
Category      &        	       Indicator            &   	   Entropy	   &    Weight \\
\hline
\multirow{3}{1.5cm}{ \centering Profitability capability }
                              &  Operating profit ratio	 &    0.586623	 &  0.062241 \\ \cline{2-4}
	                          & Return on assets&	0.661265&	0.070160\\ \cline{2-4}
	                          & Return on invested capital	&0.519275	&0.055095\\ \hline
\multirow{3}{1.5cm}{ \centering Solvency}
        	                  & Debt coverage ratio	&0.571995&	0.060689\\ \cline{2-4}
	                          & Current ratio	&0.513314&	0.054463\\ \cline{2-4}
	                          & Operating cash flow to operating profit ratio&	0.621726 & 0.065965\\ \cline{2-4}
	                          & Debt asset ratio	 &  0.648959&	0.068854\\ \hline
\multirow{3}{1.5cm}{\centering Capacity for sustainable development}
                              & Sustainable growth rate&	0.646658	&0.068610\\ \cline{2-4}
	                          & Hedging and proliferating ratios	&0.593187	&0.062937\\ \cline{2-4}
	                          & Total assets growth rate	&0.682999	&0.072466\\ \cline{2-4}
	                          & Revenue growth rate	&0.377522&	0.040055\\ \cline{2-4}
	                          & Net profit growth rate	&0.437174&	0.046384\\ \hline
\multirow{3}{1.5cm}{\centering Operation capacity	          }
                              &   Receivables turnover	&0.275741	&0.029256\\ \cline{2-4}
	                          &     Inventory turnover	&0.410767	&0.043582\\ \cline{2-4}
	                          &     Total assets turnover	&0.673957&	0.071507\\ \cline{2-4}
	                          &     Rate of cost-profit	&0.476807	&0.050589\\ \cline{2-4}
	                          &     Capital intensity	&0.727118	&0.077147\\ \hline

  \end{tabular}
\end{table}

In the evaluation of the real estate company's financial competitiveness, capital intensity is a very important indicator, its entropy is the largest in all the indicators to 0.727118, and the weight reaches 0.077147. In addition, we find that the growth rate of total assets, return on total assets and total asset turnover are with larger indicator weights, respectively, for 0.072466, 0.070160, 0.071507. This shows that the real estate company's financial competitiveness is closely related to the capabilities of asset management. As the pre-sale is the main way of selling in real estate company, accounts receivable and its management has little effect on competitiveness. The weights of receivable turnover is 0.29256, which is a minimum weight.

\subsection{Evaluation Results}
According to the formula (\ref{equ:five}), the comprehensive  score of each sample company's financial competitiveness is calculated. The scores calculated above are between 0 and 1,which make people uncomfortable. Score is 100 times magnified, then the score is between 0 to 100. Table \ref{table:top20} shows the situation of the financial competitiveness of the top 20 companies.

\begin{table}
  \centering
  \caption{the top 20 companies of the financial competitiveness}\label{table:top20}
 \noindent \begin{tabular}{|@{\;}C{0.8cm}|@{\,}C{1cm}|@{\,}C{1.6cm}|@{\;}C{0.8cm}|@{\,}C{1cm}|@{\,}C{1.6cm}|}
    \hline
 Ranking &Company &Comprehensive& Ranking	&Company &	Comprehensive \\
        &  Code    &  Score         &         & Code       &  Score         \\
        \hline

1	&000979&	69.05& 	11&	600193&	58.03\\ \hline
2	&000732&	68.03& 	12&	600791&	55.82\\ \hline
3	&000671&	67.79& 	13&	002244&	55.71\\ \hline
4	&600173&	63.21& 	14&	600684&	51.42\\ \hline
5	&600503&	63.08& 	15&	000918&	51.21\\ \hline
6	&002285&	62.87& 	16&	000506&	51.12\\ \hline
7	&600615&	60.08& 	17&	600823&	50.98\\ \hline
8	&000638&	59.29& 	18&	000558&	50.39\\ \hline
9	&000567&	59.19& 	19&	000668&	50.02\\ \hline
10	&000036&	58.51& 	20&	002016&	49.20\\

    \hline
  \end{tabular}
\end{table}

Table \ref{table:sum_of_score} shows the statistical results of the comprehensive score of the sample companies. Average score of the sample companies is 40.63, the highest score is 69.04, the lowest is 13.693. The first is  company 000979, its return on assets, capital intensity, total asset turnover, sustainable growth rate, total assets growth rate, net profit growth rate and revenue growth rate have reached 1.The second is 000732, its returns on invested capital, hedging and proliferating ratios, total assets growth rate, revenue growth rate have reached 1, the capital intensity is 0.9943, close to 1.The third is 000671, its return on assets, return on invested capital, hedging and proliferating ratios, total assets growth rate and total asset turnover are equal to 1, and its capital intensity index with the value of 0.9988 is close to 1. The company with the lowest score, its return on assets ,returns on invested capital, sustainable growth rate, operating cash flow to operating profit ratio are equal to 0,with capital-intensive index value of 0.4630. The second to last, return on assets scores 0.0687, \enlargethispage{-1.5in}return on invested capital scores 0.0475, sustainable growth rate and operating cash flow to operating profit ratio are 0, the capital intensity scores 0.1954. From the scoring results, high-scoring company has strong profitability, operational capability and high-growth, low-scoring company has weak profitability and poor ability of sustainable development. Solvency doesn't affect the company's financial competitiveness obviously.

\begin{table}[!t]
  \centering
  \caption{Descriptive Statistics of Score}\label{table:sum_of_score}
  \begin{tabular}{|c|r|}
  \hline
   Mean	&40.63030333 \\ \hline
median	&41.50017269\\  \hline
Std. Dev&	11.94833099\\  \hline
Kurtosis&	-0.02272025\\  \hline
Skewness&	0.091317864\\  \hline
Smallest&	13.69 \\  \hline
Largest	&   69.05 \\  \hline
Obs	&105 \\ \hline
  \end{tabular}
\end{table}

\section{Conclusion}
Research on financial competitiveness is still in the exploratory stage. Literature in this field is still relatively small. Companies in different industries are manifestations of differences in financial competitiveness. It is difficult to establish a evaluation system for all companies. This article established an comprehensive index competitiveness evaluation on the real estate company. According to the conceptual framework of the financial competitiveness and characteristics of the real estate company, the system consists of four parts: profitability, solvency, sustainable development and operational capacity including 17 indicators. In accordance with the principle of the information entropy, this paper proposed an objective approach to determine the indicator weight. 105 Chinese A shares listed real estate companies are evaluated. The results show that high-scoring company has strong profitability, operational capability and high-growth; low-scoring company has weak profitability and poor ability of sustainable development; Solvency doesn't affect the company's financial competitiveness obviously.


\end{document}